\documentclass[12pt,preprint]{aastex}

\def\rxte{\textit{RXTE}}
\def\msun{\mbox{M}_{\odot}}

\shorttitle{Variability of 3C~120}
\shortauthors{Marshall et al.}

\begin{document}

\title{Multi-wavelength Variability of the Broad Line Radio Galaxy 3C~120}
\author{Kevin Marshall\altaffilmark{1}\altaffilmark{2}, Wesley T.~Ryle\altaffilmark{2}, H.~Richard Miller\altaffilmark{2}, Alan P.~Marscher\altaffilmark{3}, Svetlana G.~Jorstad\altaffilmark{3}, Benjamin~Chicka\altaffilmark{4}, Ian M.~M$^{\rm c}$Hardy\altaffilmark{5}}
\altaffiltext{1}{Department of Physics, Bucknell University, Lewisburg PA 17837, {\tt kevin.marshall@bucknell.edu}}
\altaffiltext{2}{Department of Physics and Astronomy, Georgia State University, Atlanta, GA 30303}
\altaffiltext{3}{Institute for Astrophysical Research, Boston University, 725 Commonwealth Avenue, Boston MA 02215}
\altaffiltext{4}{School of Theology, Boston University, 745 Commonwealth Ave., Boston, MA 02215}
\altaffiltext{5}{School of Physics and Astronomy, University of Southampton, Southampton SO17 1BJ, UK}

\begin{abstract}
We present results from a multi-year monitoring campaign of the broad-line
radio galaxy 3C~120, using the {\it Rossi X-ray Timing Explorer (RXTE)} for
nearly five years of observations.  Additionally, we present coincident optical
monitoring using data from several ground-based observatories.  Both the X-ray
and optical emission are highly variable and appear to be strongly correlated,
with the X-ray emission leading the optical by 28~days.  The X-ray power
density spectrum is best fit by a broken power law, with a low-frequency slope
of $-1.2$, breaking to a high-frequency slope of $-2.1$, and a break frequency
of $\log \nu_{\rm b}=-5.75$~Hz, or 6.5~days.  This value agrees well with the
value expected based on 3C~120's mass and accretion rate.  We find no evidence
for a second break in the power spectrum.  Combined with a moderately soft
X-ray spectrum ($\Gamma=1.8$) and a moderately high accretion rate
($\dot{m}/\dot{m}_{\rm Edd} \sim 0.3$), this indicates that 3C~120 fits in well with the
high/soft variability state found in most other AGNs.  Previous studies
have shown that the spectrum has a strong Fe~K$\alpha$ line, which may be relativistically broadened.  The presence of this line, combined with a power
spectrum similar to that seen in Seyfert galaxies, suggests that the majority
of the X-ray emission in this object arises in or near the disk, and not in
the jet.
\end{abstract}


\section{Introduction}

Variability has long been recognized as a universal property of active galactic
nuclei (AGN).  Produced close to the nucleus, X-ray emission is often the most
variable of any waveband.  Early analysis of X-ray variability showed that the
power density spectrum (PDS) follows a power law, or ``red noise'' form,
$P \propto \nu^{\alpha}$, where $P$ is the variability power at frequency
$\nu$, and $\alpha$ is the power law slope \citep{McHardy1987,Lawrence1993}.  
No break frequency was seen, due to the short nature of the observations.  Later results demonstrated that most
AGN can be modeled by a broken power law PDS, where the slope changes from being
relatively shallow at low frequencies to some steeper value above a critical
break frequency \citep{Edelson99,Uttley2002,Markowitz2003a}.

This behavior closely mirrors what is seen in galactic black hole X-ray
binaries (BHXRBs), which contain black holes with masses of only a few $\msun$.
Much of the work connecting AGN and BHXRBs has focused on Cygnus~X-1, which is
one of the brightest and most well-studied BHXRBs.  Cygnus~X-1 is commonly in one of two
different variability states: a low/hard state with a low flux level, hard
X-ray spectrum, and a PDS with 2 separate breaks; and a high/soft state, with
higher X-ray flux, soft (steep) X-ray spectrum, and a singly-broken PDS \citep{Revnivtsev2000,McHardy2004}.
Based on scaling of the break frequencies with black hole mass and accretion rate, nearly all of the AGN studied to date match best with the high/soft state,
showing only one break in their power spectrum \citep[Akn~564 is the sole exception, see][]{McHardy2007}

The break frequency scales over more than 6 orders of magnitude in mass, if
one extrapolates the break frequencies of Cygnus~X-1 \citep{Markowitz2003a}.  The variability state
is thought to be driven by the accretion rate, with the high/soft state
correlating to a higher accretion rate which may push in the inner edge of
the accretion disk \citep{Esin1997,Churazov2001}.  This has been combined into a `fundamental plane' of
variability by \citet{McHardy2006}, with mass and bolometric luminosity
uniquely determining the break timescale.

In Seyfert~1 galaxies, where no jet is detected, the X-ray emission is mainly
thought to arise in or near the accretion disk \citep{Haardt1991,Haardt1993}.  This variability may also
be connected to the broad line region (BLR), as \citet{McHardy2006} showed
that the break timescale is also related to the full-width half-maximum (FWHM)
of H$\beta$ emission.  Thus far, efforts to analyze the X-ray variability of
AGN with jets have been limited.  Furthermore, the connection between the X-ray
variability and the jet in AGN needs to be defined better than has been possible
thus far.

The nearby ($z=0.033$) broad-line radio galaxy (BLRG) 3C~120 is an interesting
target for this type of variability analysis.  Originally classified as a
type~1 Seyfert, 3C~120 was first discovered as an X-ray source by
\citet{Forman1978} using the {\it Uhuru} observatory.  A superluminal radio jet
was later discovered by \citet{Walker1982}.  3C~120 has also exhibited
extreme optical variability at times \citep{Webb1988,French1980}, with
variations of more than half a magnitude on timescales of a few days.
\citet{Marscher2002} used coordinated {\it VLBI}\/ and {\it RXTE}\/ observations
to show that dips in the X-ray flux corresponded with ejections of superluminal
knots of bright emission along the jet.

The structure of our paper is as follows: in \S2 we discuss our observations
and data reduction methods, discuss the stationarity of the light curve in
\S3, present results for X-ray and optical correlation in \S4, present
results from Monte Carlo simulations of the power spectrum in \S5, and in \S6
we discuss implications of our results and present conclusions.

\section{Observations and Data Reduction}

We have observed 3C~120 for several years at both optical and X-ray wavelengths,
on a variety of timescales.  In Table~1 we summarize our monitoring at
each timescale.  Below we detail our data reduction methods.

\subsection{X-ray Data}

All of our X-ray observations come from a multi-year monitoring program using
\rxte.  3C~120 was monitored three times per week from 2002 March until 2007 January,
excluding several 8-week periods when the object's proximity to the Sun prevented
observations.  We refer to these data as the `long term' light curve.

Additionally, from 14 November 2006 until 12 January 2007 3C~120 was observed
every 6 hours during this period.  These data are referred to as the
`intermediate' sampling data.  Finally, 3C~120 was monitored nearly continuously
from 13--22 December 2002, which we refer to as the `short' term light curve.
The short term light curve was binned to an interval of 4000s, because AGN
typically do not show variability on shorter timescales, and such binning also
improves computational timescales greatly.

We use only data taken by PCU detectors 0 and 2, in STANDARD2 data mode.  All of our
data were reduced using FTOOLS~v5.2 software, provided by HEASARC.  Data were
excluded if the Earth elevation angle was $< 15^{\circ}$, pointing offset
$> 0.02^{\circ}$, time since South Atlantic Anomaly passage $< 30$ minutes,
or electron noise $> 0.1$ units.  Counts were extracted from the top PCU
layer only to maximize the signal to noise ratio.

Figure~1 shows the light curves for all three timescales.  In all subsequent
analysis, we do not interpolate any gaps in the light curves with the exception
of the short term light curve, which has a larger number of gaps due to Earth
occultations and other observing programs, and a lower
variance than the other data sets.  The 2--10~keV X-ray spectral index remained relatively constant throughout our observations, with all ObsIDs showing an average $\Gamma_{\rm X}=1.8$ with a
standard deviation of 0.1, where  $\nu F_{\rm x} \propto \nu^{-\Gamma_{\rm X}}$.

\subsection{Optical Data}

We have also monitored 3C~120 at optical wavelengths during most of our long
term X-ray observations using multiple ground-based observatories.  From 2004
September until 2007 January, optical data were taken using the 2~m Liverpool
Telescope at La Palma, Canary Islands, Spain.  We also supplement these
observations using data from the 1.8~m Perkins Reflector at Lowell Observatory
in Arizona,
and the 1.3~m SMARTS consortium telescope at Cerro Tololo Inter-American
Observatory in Chile.  Data were reduced using standard methods, and a
correction applied, if necessary, to match the flux measured by the Liverpool
Telescope.

All of our optical data were binned to the same 2.33~day timescale as our long
term X-ray light curve, giving us a total of 124 data points.  Figure~2 shows
the binned optical light curve, along with the relevant portion of the X-ray
light curve.

\section{Stationarity}

Before we can undertake any meaningful analysis of variability, we must first
determine whether or not the data are stationary.  In a statistical sense, 
stationarity means the underlying variability process does not change over
time.  We can test for this using the ``$S$'' statistic of
\citet{Papadakis1995}, which measures stationarity in the power spectrum.
The light curve is divided into two equal halves, the power spectrum is
computed for each half, and then the two are compared using the $S$
statistic, where $S$ is defined as:

\[ S(\nu) = \frac{\log P_1(\nu) - \log P_2(\nu)}{\sqrt{{\rm var} [
\log P_1(\nu)] + {\rm var} [\log P_2(\nu)]}} \]

\noindent where $P_{\rm 1,2}(\nu)$ are the power spectra for each half of the
light curve, and ${\rm var} [\log P_{\rm 1,2}]=0.31$ \citep{Papadakis1993}.
We can then sum over all frequencies to find the test statistic $S$,

\[ S=\frac{1}{N} \sum_{n=1}^{n_{max}} S(\nu_n) \]

\noindent where $N$ is the total number of frequencies for which we have
calculated the power spectrum.  For two identical power spectra, $S$ will have
a Gaussian distribution with zero mean and unit variance.  Or in other words,
if $|S| < 1$ we can reasonably assume the variability is stationary.

For our long-term light curve, we find that $S = 0.238$.  This indicates that
the variability in 3C~120 is stationary, and that the fundamental variability
mechanism does not change during the 4+ years of our observations.

\section{X-ray/Optical Cross Correlation}

To examine the relationship between the X-ray and optical emission, we use the
cross correlation function (CCF).  The traditional CCF requires evenly sampled
data, and can be computationally intensive.  Because of observing constraints, 
neither our X-ray nor optical data are evenly sampled.  To solve this issue,
we use the discrete correlation function (DCF) of \citet{Edelson1988}, which allows for
cross correlation of two unevenly sampled data sets.

The DCF is shown in Figure~3, with positive lag indicating X-ray emission
leading the optical.  Because the DCF has a broad shape, determining a precise
maximum value may not be the best way to quantify any possible lag, so we fit
the DCF with a Gaussian curve.  The curve provides an excellent fit to the DCF,
and has a maximum at a peak lag of +28~days, at a correlation coefficient of
0.8.

As discussed in \citet{Uttley2003}, traditional error bars are inadequate for
assessing the significance of the DCF, because adjacent data points in the
light curve are ``red noise'' data and not uncorrelated.  Therefore we use
Monte Carlo simulations to assess the significance of the correlation.  Our
method is as follows.

We began by simulating 2 independent red noise light curves, using the method
of \citet{Timmer1995}.  One light curve was given a power spectrum identical to
the X-ray PDS in \S5, and the other light curve was simulated using the
measured optical PDS, which shows a slope of $-1.6$ with no measured break
(Ryle et al.~2009, in preparation).  The two light
curves were then re-sampled in the same fashion as the original light curves, 
and random noise was added in the form of a Gaussian random with mean of zero
and standard deviation equal to the average observed error.  The light curves
were then correlated, looking for any cases where the maximum correlation
coefficient was greater than 0.8.  Out of 1000 simulations, this was only the
case 2 times.  Therefore the correlation coefficient of $r=0.8$ seen in the
data is significant at more than 99\% confidence.

To explore the significance of the measured lag, we use the Flux Randomization
/ Random Subset Selection (FR/RSS) technique \citep{White1994,Peterson1998}.
Briefly, random errors are added to each light curve, using the same method as
above.  For each light curve with $N$ data points total, the same number of
data points are chosen at random from each curve, and duplicates are rejected.
This typically results in a fraction of $\sim 1/e$ points rejected from each new light curve.  The curves are
then cross correlated as before, the DCF is fit with a Gaussian, and the centroid noted.  This is
done a large number of times (1000 simulations), and the distribution of
DCF centroid values is used to estimate uncertainties.

As noted by \citet{Maoz1989}, distributions of lags are
nonnormal and the standard deviation is not a good estimate of the uncertainty
in the lag.  Therefore similar to \citet{Peterson1998}, we quote uncertainties
at $\pm \Delta \tau_{68}$, where 68.27\% of the realizations yield results
between $\tau_{\rm median}-\Delta \tau_{68}$ and $\tau_{\rm median}+\Delta \tau_{68}$, which corresponds to a $1 \sigma$ error for a normal distribution.
Using this method, we estimate the lag for 3C~120 as $28.73_{-5.87}^{+6.19}$~days.  The distribution of lags measured using the FR/RSS method is shown in
Figure~4.

\section{X-ray Power Density Spectrum}

The large amount of X-ray data, covering a wide variety of timescales, allow us
to calculate the power density spectrum (PDS) over more than 4 decades of
frequency.  Calculating the PDS for any discretely sampled, finite length data
set is inherently problematic, suffering from a number of complications such as
windowing, aliasing, and red noise leakage.  To avoid these complications, we
follow the method of \citet{Uttley2002}, which we outline briefly below.

We begin by calculating the square modulus of the discrete Fourier transform
for each light curve, using the formula

\[ |F(\nu)|^2 = \left[ \sum_{i=1}^{N} f(t_i) \cos(2 \pi \nu t_i) \right]^2 + \left[ \sum_{i=1}^{N} f(t_i) \sin (2 \pi \nu t_i) \right]^2, \]

\noindent where $t_i$ are the light curve data points, and $|F(\nu)|^2$ is
calculated at evenly spaced frequencies from $\nu_{\rm min}=1/T$ to $\nu_{\rm Nyq}=N/2T$, where $T$ is the total duration of the light curve.  The PDS is then
normalized using the fractional rms squared normalization \citep{Miyamoto1991},

\[ P(\nu) = \frac{2T}{\mu^2 N^2} |F(\nu)|^2, \]

\noindent where $\mu$ is the average count rate of the light curve.  The
observed PDS is then binned geometrically every factor of 1.5 (0.18 in the
logarithm).

Monte Carlo simulations are then used to find the best-fit model for the
observed PDS.  A simulated light curve is generated based on an input model
power spectrum, based on the method of \citet{Timmer1995}.  The simulated light
curve is created with at least 100 times the duration of the actual light curve,
to avoid the problem of variability power from longer timescales being shifted
to shorter timescales (red noise leakage).  The simulated light curve is then
split into 100 segments of equal length.

Because our data are not continuously sampled, power above the Nyquist frequency
can also be aliased back into the power spectrum.  To avoid this, our simulated
light curves have a time resolution of $0.1 T_{\rm samp}$, where $T_{\rm samp}$
is the sampling timescale of the light curve.

The simulated light curves are then re-sampled in the same fashion as the
observed light curves.  The PDS is then calculated for each of the 100 curves,
and binned by factors of 1.5.  The 100 individual power spectra are then
averaged, and error bars for each data point are calculated from the rms spread
of the individual PDS points.

Poisson noise is added to the simulated power spectra using

\[ P_{\rm Poisson} = \frac{\sum_{i=1}^{N} \sigma(i)^2}{N(\nu_{\rm Nyq} - \nu_{\rm min})}, \]

\noindent where $\sigma(i)$ are the error bars for each point in the light curve.  We can also estimate analytically the amount of power which will be aliased
back into the PDS from frequencies above the minimum resolution of the simulations.  This power is given by:

\[ P_{\rm Alias} = \frac{1}{\nu_{\rm Nyq} - \nu_{\rm min}} \int_{\nu_{\rm Nyq}}^{1/(2T_{\rm exp})} P(\nu) d\nu, \]

\noindent where $T_{\rm exp}$ is the exposure time of our observations,
typically $\sim 1$~ksec.

To determine the goodness of fit of the model, we use a modified $\chi^2$ fit.
First we compare the observed PDS to average of the simulated power spectra:

\[ \chi_{\rm obs}^2 = \sum_{\nu} \frac{(\overline{P_{\rm sim}(\nu)} - P_{\rm obs}(\nu))^2}{(\Delta \overline{P_{\rm sim}(\nu)})^2}, \]

\noindent where $P_{\rm obs}(\nu)$ is the observed PDS at frequency $\nu$, $\overline{P_{\rm sim}(\nu)}$ is the average of the simulated power spectra at that
same frequency, and $\Delta \overline{P_{\rm sim}(\nu)}$ is the rms spread of
the power spectra.

We then randomly choose 1000 combinations of simulated power spectra at each
sampling rate.  We then calculate $\chi_{\rm dist}^2$, where:

\[ \chi_{\rm dist}^2 = \sum_{\nu} \frac{(P_{\rm sim}(\nu)-\overline{P_{\rm sim}}(\nu))^2}{\Delta \overline{P_{\rm sim}}(\nu)^2}. \]

Of these 1000 combinations, the fraction where $\chi_{\rm obs}$ is smaller than
$\chi_{\rm dist}$ represents the probability of acceptance of the model,
i.e., the probability that the model provides an acceptable fit to the data.
This process is repeated for a grid of slopes and break frequencies, to
determine the parameters which yield the highest probability of acceptance.

\subsection{Unbroken Power Law}

We began by fitting the observed PDS with an unbroken power law of the form
$P(\nu) \propto \nu^{\alpha}$.  This yields a best-fit power law slope of
$\alpha=-1.5$, with a probability of acceptance of only 1.3\%, and is clearly
not an adequate fit to the data.

\subsection{Broken Power Law}

Visual inspection of the PDS indicates that the shape may be a broken power law,
with a fixed slope at low frequencies that breaks to a steeper slope at high
frequencies.  For our simulations, we use a broken power law of the form:

\[ P(\nu) = A \left( \frac{\nu}{\nu_{\rm b}} \right) ^{\alpha_{\rm L,H}}\]

\noindent where A is the normalization, $\nu_{\rm b}$ is the break frequency, $\alpha_{\rm L}$ is the low-frequency
slope, and $\alpha_{H}$ is the high-frequency slope. The low frequency slope
was allowed to vary from $-1.0$ to $-1.5$ in increments of 0.1, and the high frequency slope was allowed to vary between
$-1.5$ and $-3.5$ in increments of 0.2.  The break frequency was allowed to
vary between $\log \nu_{\rm b}=-8.0$ and $-4.0$, incrementing the break by
factors of 2 (0.30 in the logarithm).  Once a fit was obtained, a finer grid
with increments of 0.1 in slope and 0.0414 in break (factors of 1.1) were used
to improve the best-fit values.

Figure~5 shows the observed PDS fitted with the best-fit broken power law model.
The best fit parameters show a low-frequency slope of $\alpha_{\rm L}=-1.2$, high-frequency slope $\alpha_{\rm H}=-2.1\pm0.4$, with a break at a frequency
of $\log \nu_{\rm b}=-5.75\pm0.43$~Hz, or a timescale of $6.51^{+11.0}_{-4.1}$~days.  The probability of
acceptance of this model is 84.5\%.  Figure~6 shows confidence contours for
variations in the break frequency and high frequency slope.  Error bars for the
slope and break were obtained by taking a 1-dimensional slice through the
confidence contours at the best-fit value, fitting the contours with a Gaussian and finding the standard deviation.

\subsection{Doubly-Broken Power Law}

Given that the low/hard state seen in BHXRBs shows two break frequencies, we
also model the PDS of 3C~120 with a doubly broken power law.  For computational
efficiency, the slopes were fixed at values of $0$, $-1$, and $-2$, similar
to that seen in Cygnus~X-1 \citep{Revnivtsev2000}.  The low frequency break
was allowed to vary from $\log \nu_{\rm b,lo}=-9.0$ to $-5.5$, and the high
frequency $\log \nu_{\rm b,hi}$ allowed to vary between 5 and 10,000 times
greater than the low-frequency break.

A best fit was obtained at $\log \nu_{\rm b,lo}=-8.03$, $\log \nu_{\rm b,hi}=-5.83$, and a confidence level of 63.2\%.  The low frequency break is near the
lowest measured frequency for our PDS, effectively turning the model into a
singly-broken power law.  Therefore we find no evidence for a doubly-broken
power law in the PDS of 3C~120, with a singly-broken power law providing an
adequate fit.

\section{Discussion and Conclusions}

We have shown that both X-ray and optical emission are highly variable over
a period of several years, with fractional variabilities of 21.6\% and 8.2\% in
the X-ray and optical, respectively \citep{Markowitz2004}.  The variations are strongly correlated, with the
X-ray variations leading the optical by 28~days.  Additionally, the X-ray
PDS is best fit by a broken power law, with a slope of $-1.2$ breaking to a
slope of $-2.1$ above a break frequency of $\nu_{\rm b}=-5.75$~Hz, or 6.5~days.

Given that the PDS is best fit by a singly-broken power law, and does not show
any signs of a double break, 3C~120 appears to fit well with the high/soft
state power spectra found in most other AGN.  This is appropriate, given the
moderately high accretion rate of 30\% Eddington for this object \citep{Ogle2005}.
The X-ray spectrum is also relatively steep, with an average photon index of
$\Gamma=1.8$ during our long-term monitoring, similar to the values of
$\Gamma=2-3$ seen in high/soft state BHXRBs.

Using the fundamental plane of AGN variability \citep{McHardy2006}, we can investigate whether
the break timescale measured for 3C~120 fits with a relation based on
radio-quiet objects.  The expected break timescale $T_{\rm B}$ is:

\[ \log T_{\rm B} = 2.1 \log M_{\rm BH} - 0.98 \log L_{\rm Bol} - 2.32 \]

\noindent where $M_{\rm BH}$ is the mass of the black hole in units of
$10^6 \msun$, and $L_{\rm Bol}$ is the bolometric luminosity in units of
$10^{44}$ erg/s.  For 3C~120, $L_{\rm Bol} = 21.87$ \citep{Woo2002}, and
$M_{\rm BH}=50^{+108}_{-18}$, as measured by \citet{Vestergaard2006} using
reverberation mapping in the optical and UV.  Using these values, we would expect to
find a break at a timescale of $T_{\rm B}=0.86^{+8.78}_{-0.52}$~days, with the
errors based on the large uncertainty in the mass.
Therefore our measured break timescale agrees well with the prediction based
on the fundamental variability plane, within the errors on both the mass and
our measured break.

We can also check the relationship between H$\beta$ FWHM and $T_{\rm B}$ for
3C~120.  This relationship is given by:

\[ \log T_{\rm B} = 4.2 \log \rm{H}_{\beta,FWHM} - 14.43 \]

\noindent where $\rm{H}_{\beta,FWHM}$ is in units of km/s \citep{McHardy2006}.  For 3C~120, \citet{Marziani2003} measured a FWHM of 2328 km/s.  This gives us a predicted break timescale of
0.51~days, different by a factor of $\sim 2$ from the mass/luminosity scaling
relationship, and different by a factor of $\sim 13$ from our actual measured
value.  The $H \beta-T_{\rm B}$ relationship is based on a sample of radio-quiet
objects.  Given that 3C~120 is radio loud, with a superluminal jet, the
broad-line region may not share the same characteristics as a radio-quiet
object.  More observations of radio-loud AGN are needed to determine if
a separate $H\beta-T_{\rm B}$ relationship exists for these objects.

Given that the X-ray behavior of this object appears similar to that from
radio-quiet Seyfert galaxies, we can examine whether the X-ray emission comes
from the jet, or from a source in/near the accretion disk.  As noted by
\citet{Ogle2005}, although 2 X-ray sources are coincident with the positions
of radio lobes, they contribute $< 5$\% of the overall X-ray flux.

Closer to
the nucleus, we can use spectroscopy to determine how much of the flux 
originates in the jet.  \citet{Ballantyne2004} have analyzed a 127~ks
{\it XMM/Newton} observation of 3C~120.  They find a neutral Fe~K$\alpha$
line with an equivalent width of 53~eV.  \citet{Kataoka2007} find evidence for
a relativistically broadened line using Suzaku.  The strength of the narrow Fe~K$\alpha$ line
indicates that the majority of the X-ray emission has its origin in an unbeamed
source, such as the accretion disk or corona, as opposed to the jet.  If the
X-ray flux were beamed within a narrow cone, reflection would be unlikely.
\citet{Ogle2005} find a half-opening angle for the continuum of $> 50^{\circ}$,
indicating the narrow jet is not responsible for the X-ray emission.

If the X-ray emission originates near the accretion disk, then where does the
28~day X-ray to optical lag arise from?  Low polarization in the optical
($\sim 0.3$\%, consistent with interstellar dust) indicates that the emission
is not associated with the jet.  Such a lag is also too large to arise from
light travel time between different regions in the accretion disk, however
the communication between X-ray and optical regions may propagate at sub-light
speeds.  A more complete optical light curve may give a more precise value of
the delay, to confirm the result and better specify the required propagation
speed of the variations.

\acknowledgements{We thank an anonymous referee for a thorough report which
improved this paper.  KM, WRT, and HRM acknowledge support from NASA grant
No. NNG04G046G, and from the PEGA program at GSU.  The research at Boston
University was supported in part by National Science Foundation grant
AST-0406865 and a number of NASA grants, most recently NNX06AH12G and
NNX08AJ64G.  The Liverpool Telescope is operated on the island of La Palma by
Liverpool John Moores University in the Spanish Observatorio del Roque
de los Muchachos of the Instituto de Astrofisica de Canarias with
financial support from the UK Science and Technology Facilities Council.}

\bibliographystyle{apj}

\bibliography{ms}

\begin{deluxetable}{ccccc}
\tablewidth{0pt}
\tablecaption{Summary of Observations}
\tabletypesize{\scriptsize}
\tablehead{
\colhead{Duration} & \colhead{Start\tablenotemark{a}} & \colhead{End\tablenotemark{a}} & \colhead{T\tablenotemark{b}} & \colhead{$\Delta$T\tablenotemark{c}}}

\startdata

Long & 52335.42 & 54114.39 & 1779 & 2.33$^{\rm d}$\\
Medium & 54054.66 & 54114.39 & 60 & 6$^{\rm h}$\\
Short & 52622.39 & 52631.48 & 9.09 & 1000$^{\rm s}$\\
Optical & 53248.73 & 54110.54 & 862 & 2.33$^{\rm d}$\\

\enddata
\tablenotetext{a}{Modified Julian Date}
\tablenotetext{b}{{}Length of light curve, in days}
\tablenotetext{c}{{}Light curve binning}

\end{deluxetable}

\begin{figure}
\figurenum{1}
\plotone{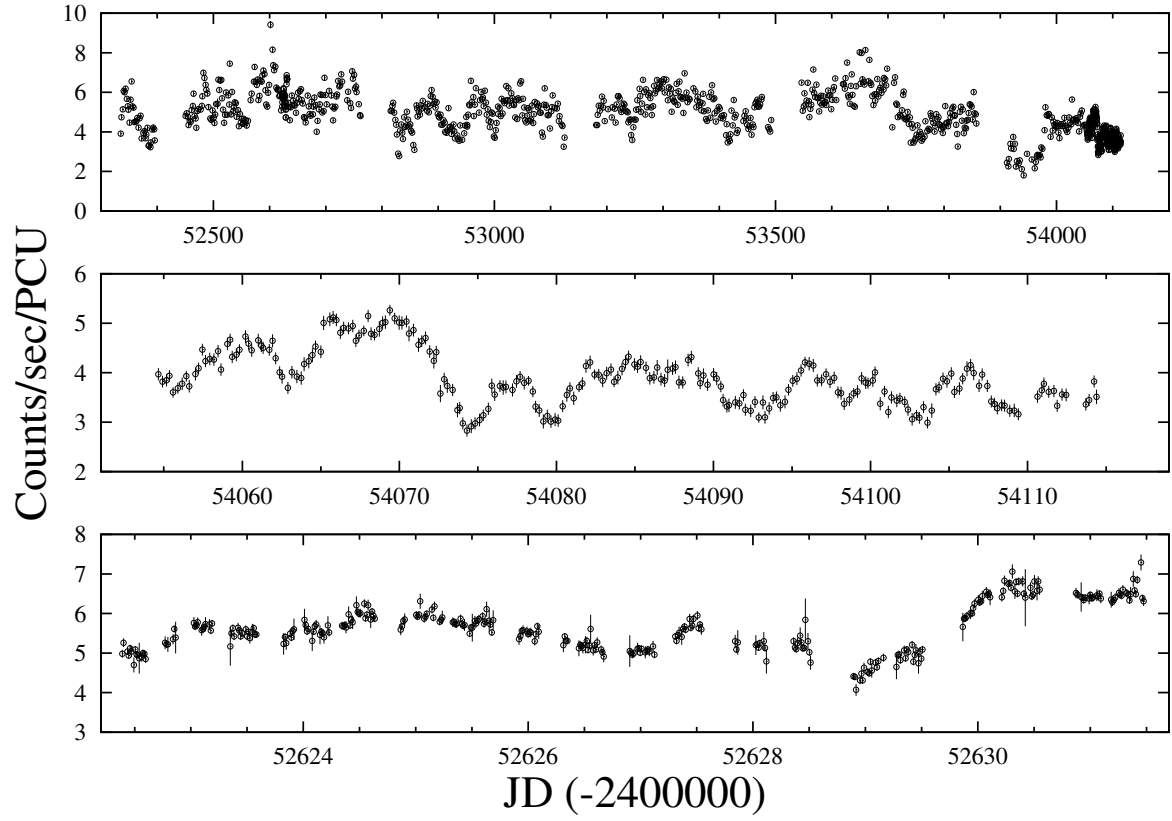}
\caption{\rxte\/ 2--20~keV light curves for long (top), medium (middle), and short (bottom) timescales.}
\end{figure}

\begin{figure}
\figurenum{2}
\plotone{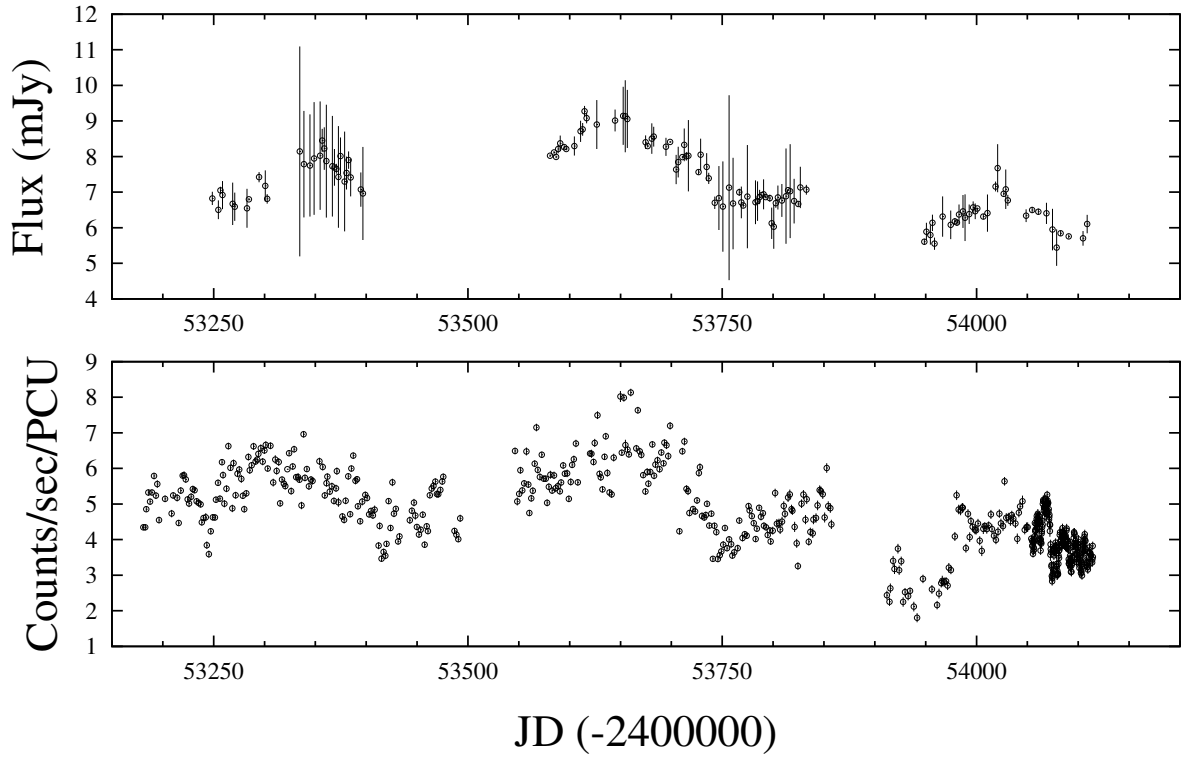}
\caption{Optical R-band (top) and X-ray (bottom) light curves.}
\end{figure}

\begin{figure}
\figurenum{3}
\plotone{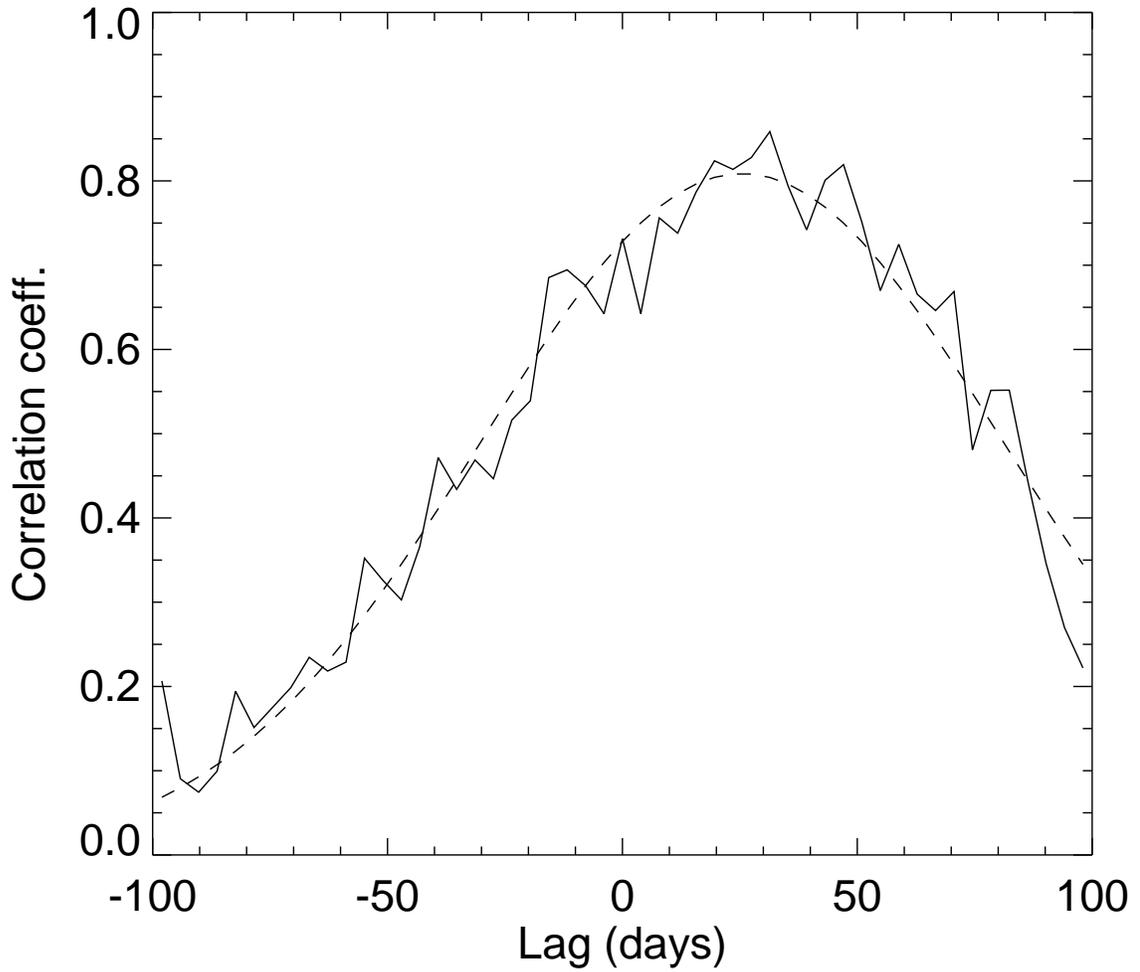}
\caption{Discrete correlation function for X-ray and optical light curves.  Positive lag indicates X-ray emission leading the optical.  Dashed line shows a best-fit Gaussian curve, with centroid of +28~days.  See text for a discussion of errors.}
\end{figure}

\begin{figure}
\figurenum{4}
\plotone{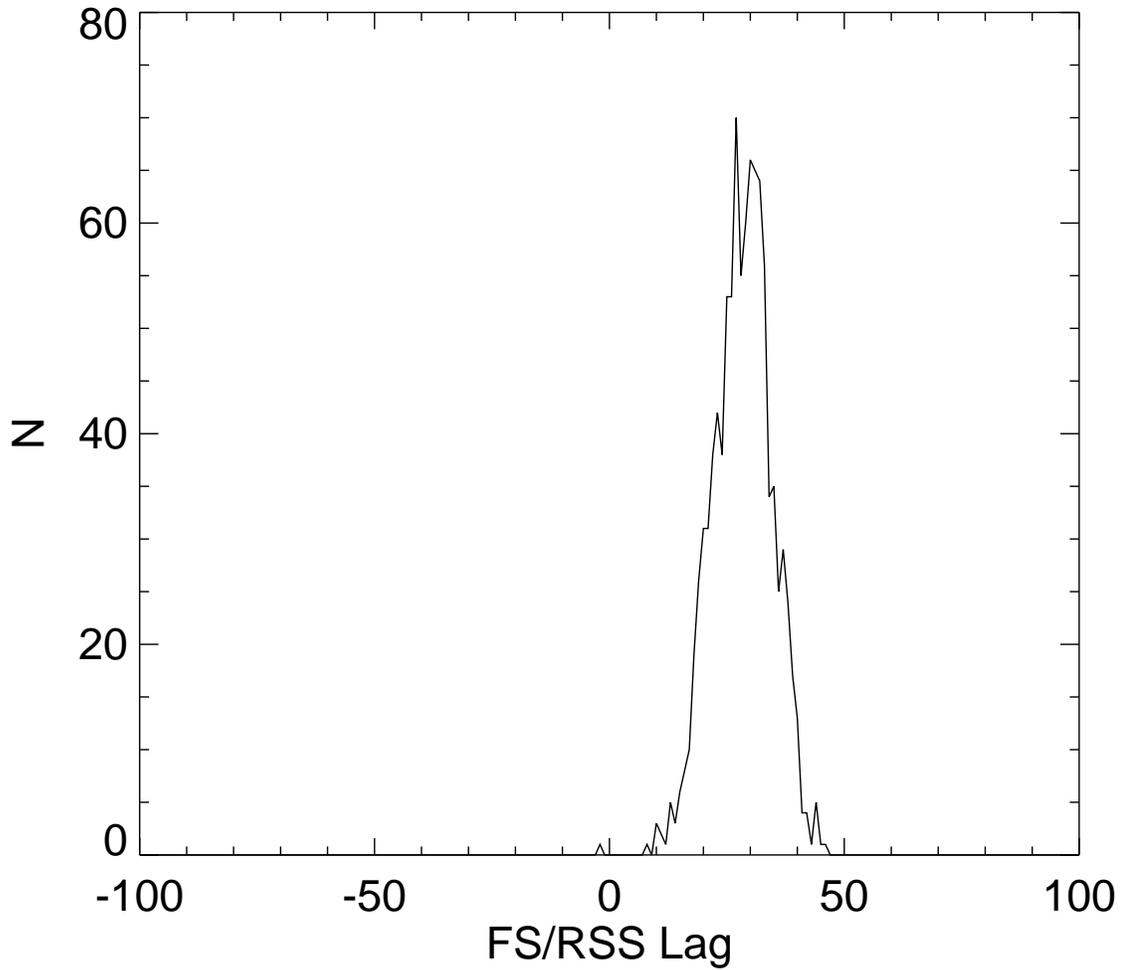}
\caption{Histogram showing cross-correlation lag centroid versus number of
simulation results yielding that lag for the FR/RSS method.  Bin width is 1~day.  Median result occurs
at a lag of 28.73 days, with uncertainties of $+\Delta\tau_{68}=6.19$ and $-\Delta\tau_{68}=5.87$ days.}
\end{figure}

\begin{figure}
\figurenum{5}
\plotone{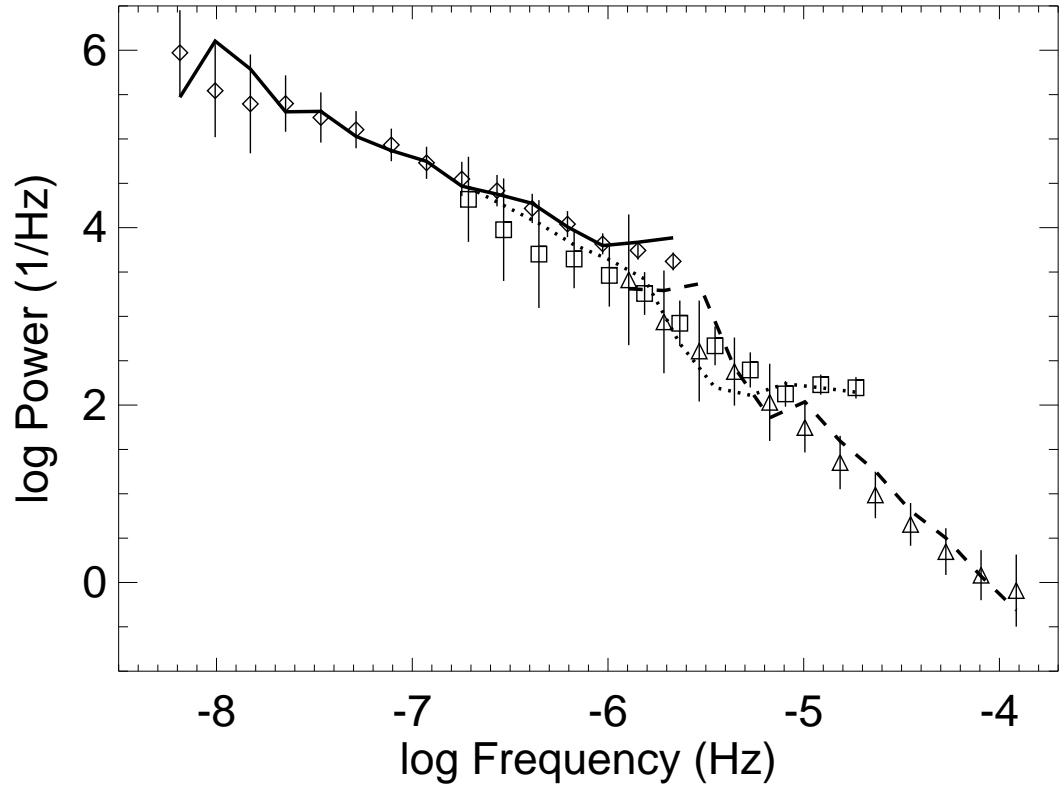}
\caption{Observed PDS for long, medium, and short light curves represented by solid, dotted, and dashed lines, respectively.  Monte Carlo simulations for each light curve are represented by diamonds, squares, and triangles with error bars.  Best fit occurs with the low frequency slope fixed at $-1.2$, high frequency slope of $-2.1$, and a break frequency of $\log \nu_{\rm b}=-5.75$~Hz, with a probability of acceptance of 84.5\%.}
\end{figure}

\begin{figure}
\figurenum{6}
\plotone{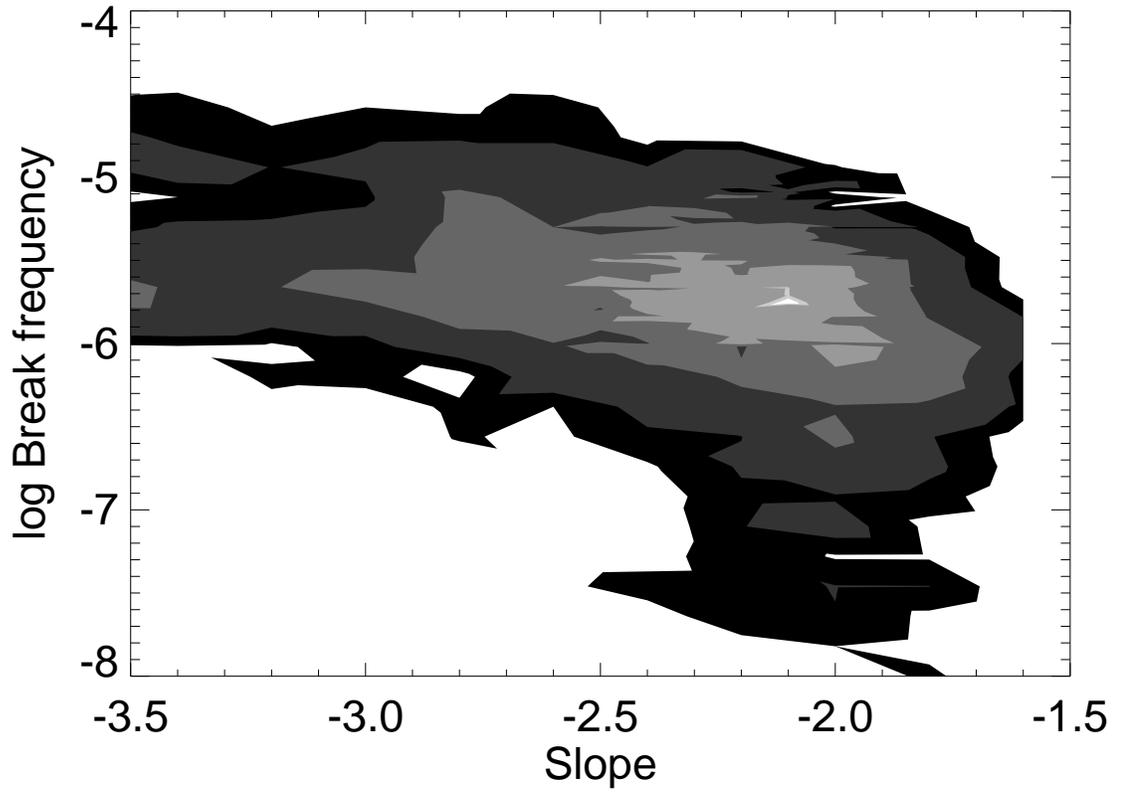}
\caption{Confidence contours for a broken power law model with low frequency slope of $-1.2$.  Contours are at 5, 10, 25, 50, and 75\% probability of acceptance.  Maximum confidence of 84.5\% occurs at a high-frequency slope of $-2.1$ and a break frequency of $\log \nu_{\rm b}=-5.75$~Hz.}
\end{figure}

\end{document}